A single step protein assay that is both detergent and reducer compatible: the cydex blue assay

Thierry Rabilloud 1, 2, 3

1: CNRS UMR 5249, Laboratory of Chemistry and Biology of Metals, Grenoble, France
2: Univ. Grenoble Alpes, Laboratory of Chemistry and Biology of Metals, Grenoble, France
3: CEA Grenoble, BIG/CBM, Laboratory of Chemistry and Biology of Metals, Grenoble, France

to whom correspondence should be addressed:
Laboratoire de Chimie et Biologie des Métaux, UMR CNRS-CEA-UJF 5249, BIG/CBM, CEA Grenoble, 17 rue des martyrs, F-38054 Grenoble Cedex 9, France
thierry.rabilloud@cea.fr

Abstract

Determination of protein concentration in samples for biochemical and proteomic analyses is often a pre-requisite. However, the protein assay methods available to date are not compatible with both reducers and detergents, which are however present simultaneously in most of the denaturing extraction buffers used in proteomics and electrophoresis, e.g. SDS electrophoresis.
It was found that inclusion of cyclodextrins in a Coomassie blue-based assay makes it compatible with detergents, as cyclodextrins complex detergents on a 1:1 molecular basis. This type of assay being intrinsically resistant to reducers, a single step assay that is both detergent and reducer compatible is obtained. Depending on the type and concentration of detergents present in the sample buffer, either beta-cyclodextrin or alpha-cyclodextrin can be used, the former being able to complex more types of detergents and the latter being able to complex higher amounts of detergents due to its greater solubility in water. Cyclodextrins are used at final concentrations of 2-10 mg/ml in the assay mix. This typically allows to measure samples containing both down to 0.1 mg/ml protein, 2% detergent and reducers (e.g. 5% mercaptoethanol or 50mM DTT) in a single step, with a simple spectrophotometric assay.

Introduction

Measurement of protein amounts is an important step in all proteomic experiments, whatever the proteomic setup used. In comparative proteomics, whether 2D gel-based or through shotgun techniques it is necessary to use very comparable protein amounts for all the samples used in the comparison. Even in the validation steps, e.g. by western blotting, normalization by the amount loaded is by far the best normalization process [1, 2]. However, even post electrophoretic normalization requires comparable amounts of proteins to be loaded on the different lanes of the gel and thus prior measurement of protein concentration in the samples of interest.

Protein concentration determination is carried out by two families of methods. in the first one, the biuret reaction is used, in which the peptide bond reduces $Cu^{2+}$ to $Cu^+$ under alkaline conditions. The $Cu^+$ ion is then used as a reducer of the Folin reagent to produce a blue-colored compound [3] or is read directly as a colored complex with bicinchoninic acid [4]. In the second family of methods, protein binding to dyes and metachromic effects of this binding are used. The most widely used method of this family is the Bradford method [5], in which Coomassie blue G is converted from its leuco, brown form into its blue form through protein binding. The pyrogallol red-molybdate [6] and pyrocatechol violet-molybdate [7] methods also belong to this family, but are more used in a specialized context (e.g. urinary protein determination).

Unfortunately, all these methods are subjects to strong interferences by chemicals that are very commonly used in protein extraction for proteomics and/or blotting. In these methods, the proteins must be fully denatured and reduced in the solubilization process to ensure an as complete and as thorough protein extraction as possible, and also to make to process highly reproducible. This requires the use of reducers (e.g. mercaptoethanol, dithiothreitol, tributyl- or tris(carboxyethyl)- phosphine and of strong protein denaturants such as urea or more commonly strong detergents such as SDS.
On the one hand, reducers induce a strong interference on the copper-based protein assays (Lowry and BCA) by reducing the $Cu^{2+}$ into $Cu^+$ without the presence of any protein. On the other hand, many surfactants bind to dyes and induce the metachromic effect required in the dye-binding methods (e.g. the Bradford method) in the absence of protein. These interferences lead to unacceptable backgrounds in the reagent blanks and preclude measurements of protein concentrations.
There are however some solutions to this problem, as reviewed recently [8] . One is to destroy the reducers used in the protein extraction solutions by alkylation, then to use a copper-based method to assay the protein concentration [9]. Other methods use precipitation of the protein sample to remove all the interfering substances, either in batch [10, 11] or on filters [12-14]. Depending on the format used, either copper-based proteoin assays [10, 11] or dye-based assays [12-14]. As an other example, a protein precipitation step is used in the commercial RC/DC protein determination kit (Bio-Rad)

Although efficient, these methods are not trouble-free. They introduce additional steps which may introduce variability in protein recovery (e.g. the precipitation step) and thus variability in the final protein assay. As to the alkylation strategy
[9], it requires a massive excess of alkylating agent over the reducing one, which is not always easy to achieve. Furthermore, very effective reducers such as phosphines do not alkylate well with iodoacetamide, and give more erratic results in this procedure. Last but not least, copper-based protein assays are two-step assays which require rather precise timing, and are thus less easy to use as a routine laboratory procedure than dye binding assays that are single-step assays.

Thus, it would be interesting to render the dye methods and especially the widely used Bradford method insensitive to detergents, as the method is intrinsically already insensitive to reducers. Without any modification, the Bradford method is insensitive to CHAPS and to low concentrations of sulfobetaine detergents. Extra acidification makes it more tolerant to nonionic detergents such as NP40/Triton X-100 [15], but is still remains very sensitive to SDS [15] and other ionic detergents.

As the basis of the detergents interference with dye-based protein assays is the bonding of the detergents molecules to the dye, a good way to avoid this interference would be to complex the detergent molecules present in the sample in order to prevent their binding to the dye molecules. Cyclodextrins are known for a long time to complex linear detergents [16] but also larger molecules such as sterols [17, 18] or polyisoprenic chains [19]. Their affinity for SDS, which is the most commonly used detergent in proteomics and in blotting, has been documented in detail [20, 21]. It was therefore decided to test whether the inclusion of cyclodextrins in a Bradford assay would make it detergent compatible and if yes, up to which extent and for which detergents. As the various cyclodextrins have different internal cavities and thus complex molecules of different geometries, the work was focused on alpha-cyclodextrin, which complexes efficiently linear molecules and should be efficient for lnear detergents such as SDS, and beta-cyclodextrin, which complexes larger molecules such as sterols, and thus should able to bind more bulky detergents such as bile salts (sterol-derived) or Triton X-100 (tert-octylphenyl-derived).

Material and methods

All the experiments were performed in duplicate. This allows to investigate the variability of the method while giving a high estimate of the variability.

Alpha and beta-cyclodextrins were purchased from Sigma or Alfa-Aesar, and dissolved in hot water prior to use. Beta-cyclodextrin can be dissolved at 5 mg/ml, and alpha-cyclodextrin at 50 mg/ml. These solutions are stable for several days at room temperature.

The protein assay was performed in a tube format. The cyclodextrin solution (up to 0.5 ml) was mixed with water and with the sample solution to reach 0.8 ml. Then 0.2 ml of concentrated Bradford reagent (Bio Rad, catalog number #500-0006) was added with a positive displacement pipette (e.g. Gilson Microman 250) in order to reduce the variability that may be brought by the viscosity of the dye concentrate. The content of the tubes were throughly mixed, the color let to develop for at least 5 minutes at room temperature, and the absorbance at 595 nm was finally read.

Bovine serum albumin (pre-made 1mg/ml solution from Sigma, catalog number P0914) was used for preliminary tests and as a standard protein.

For complex samples testing, RAW264 cells were grown in RPMI1640 medium supplemented with 10% fetal bovine serum. The cells were harvested by scraping in the culture medium, then collected by centrifugation (100g, 5 minutes), and rinsed three times with PBS. The volume of the cell pellet was then estimated, and the protein extracted by ten pellet volumes of extraction solution. Several extraction solutions were tested:

i) 7M urea, 2M thiourea, 4% CHAPS, 30mM Spermine base, 60mM DTT. After extraction at room temperature for 1 hour, the extract was centrifuged at 200,000g for 30 minutes at 20°C, and the supernatant collected.

ii) 10mM Hepes-NaOH pH 7.5, 2mM MgCl2, 50mM KCl, 0.1% 3-[tetradecyldimethylammonio]-propane-1 sulfonate (SB 3-14). After extraction for 30 minutes on ice, the extract was centrifuged at 15,000g for 15 minutes at 4°C, and the supernatant collected.

iii) RIPA buffer (from Cell Signaling technology). The buffer contains 20mM Tris-HCl pH 7.5,

150mM NaCl, 1mM EDTA, 1mM EGTA, 1% NP-40, 1% sodium deoxycholate, 2.5 mM sodium pyrophosphate, 1mM glycerophosphate, 1mM sodium vanadate and 1μg/ml leupeptin. The extraction was carried out as for solution ii above.

iv) 100mM Tris-HCl pH 7.5, 2% SDS and reducers. Three different reducers were tested: either 5% (v/v) mercaptoethanol, or 50mM DTT or 10 mM TCEP. The samples were extracted and denatured at 70°C for 30 minutes, then centrifuged at 15,000g for 15 minutes at room temperature. The viscous supernatant was collected and sheared through repeated passage in a 0.8 mm diameter syringe needle.

v) 100mM pyruvic acid, 50mM NaOH, 10mM TCEP and 2% SDS. The samples were extracted and denatured at 70°C for 30 minutes, then centrifuged at 15,000g for 15 minutes at room temperature. A massive white pellet was obtained, and the fluid supernatant was collected.

vi) 100mM pyruvic acid, 50mM NaOH, 10mM TCEP and 2% CTAB. The samples were extracted and denatured at 70°C for 30 minutes, then centrifuged at 15,000g for 15 minutes at room temperature. A massive white pellet was obtained, and the fluid supernatant was collected.

For protein assays of the complex samples, a small aliquot of each sample was diluted 10 times in its own extraction buffer. 10 μl of this dilution were used as a sample in the protein assay. For each extract, a standard curve was build with variable amounts of BSA, mixed with 10 μl of the corresponding extraction buffer.

In order to check the results of the protein assays, SDS electrophoresis was performed. The samples obtained from RAW 264 cells were diluted at 1mg/ml (as derived from the protein assays) in SDS sample buffer (100mM Tris-HCl pH 7.5, 2% SDS, 50mM DTT, 20% glycerol and a trace of bromophenol blue). Samples extracted under conditions ii and iii were then denatured at 70°C fro 30 minutes prior to use. 40μg of proteins were loaded on top of a large (160x200x1.5 mm) 10% acrylamide gel, operated in the tris-taurine system [22]. The empty wells separating the protein-loaded wells were loaded with equal volumes (40μl) of SDS sample buffer. After electrophoresis at 12W/gel and 10°C, the gels were stained with colloidal Coomassie Blue [23].

Results

As a first test, the ability of cyclodextrin to induce tolerance of the Bradford protein assay to detergents was investigated. Three commonly-used detergents (SDS, Triton X-100 and CTAB) were tested, and the results are shown on Figure 1. Without cyclodextrins, the assay was extremely sensitive to ionic detergents, with unacceptable backgrounds obtained at 0.1 mg detergents in the final protein assay volume (1ml) , i.e. a 5µl volume of a sample solution containing 2% detergent. 0.1 mg Triton X-100 was still tolerable, but 0.2 mg was not (Figure 1C). When cyclodextrins were added to the assay, several phenomena were observed. First the addition of beta-cyclodextrin induced an increase in the absorbance of the blank (without detergent) compared to the control without cyclodextrins added, while alpha-cyclodextrin did not induce such an increase in absorbance. Second the absorbance in the presence of beta-cyclodextrin decreased when detergent was added. Third and more importantly, the cyclodextrins were able to bind detergent and prevent the detergent induced interference. This binding was concentration dependent and occurred up to equivalence point (1 detergent molecule per cyclodextrin molecule) Above this equivalence point, the detergent interference resumed again. Fourth, in this landscape, alpha-cyclodextrin did not complex Triton X-100 well, while beta-cyclodextrin did (Figure 1C). Conversely, beta-cyclodextrin did also complex linear detergent well, but is less efficient than alpha-cyclodextrin just because less amounts can be used.

The response curve of the assay in the presence of cyclodextrins was then investigated, and the results are shown on Figure 2. The high blank absorbance induced by beta-cyclodextrin can be seen here again, and it can also be seen that cyclodextrins shallowed the slope of the dose response curve in a dose-dependent manner. The mean slope was of 60mOD/µg albumin for the control, 45 mOD/µg albumin for the 5mg/ml alpha-cyclodextrin condition, 38 mOD/µg albumin for the 10mg/ml alpha-cyclodextrin condition and 22 mOD/µg albumin for the 2. 5mg/ml beta-cyclodextrin condition.

These two results were then combined to test a protein assay with both detergents, cyclodextrins and BSA. The first detergent to be tested was SDS, as it is the most widely used. The results for a final concentration of 5mg/ml alpha-cyclodextrin in the assay are shown on Figure 3A. They demonstrate that the assay is insensitive to up to 30µl of SDS containing buffer (Tris 100mM pH 7.5, 2% SDS, 5% mercaptoethanol), and becomes erratic at 40µl of SDS buffer per ml of final assay. The influence of the reducer present in the SDS buffer was then tested, and the results are shown on Figure 3B. It can be seen that at 20µl of SDS containing buffer per ml of assay, the reducer (either 5% mercaptoethanol, 50mM DTT or 10mM TCEP) has a minimal influence on the protein assay. Furthermore, the packing of the curves on Figure 3A shows that the response of the assay does not change much (less than 10%) in the 0-20µl range of sample buffer. This further shows the robustness of the assay

The opposite class of ionic detergents, namely cationic detergents, was then tested. Cationic detergents are used in some specialized protocols using then as the first dimension of two-dimensional gels [24] or as an alternative to SDS gels in some cases [25] . Cationic detergents induce however an even more severe interference than anionic ones, as the binding of detergents to Coomassie blue is favored by the opposite charges of the two molecules. The results, shown on Figure 4 A, demonstrate that a useful proteins assay can be devised using 5mg/ml alpha-cyclodextrin, tolerating up to 25µl of CTAB-containing buffer (Pyruvic acid 100mM, NaOH 50mM, CTAB 2%, TCEP 10mM). The amount of tolerable CTAB buffer can be increased to up to 50µl when the alpha-cyclodextrin concentration is increased to 10 mg/ml (Figure 4B). Here again, the response of the assay changes by less that 10% at the lower end of the curve, and by less than 5% at the upper hand (Figure 4A) showing the robustness of the assay for the 0-20µl range of sample buffer.

To complete these tests with detergents, a complex detergent-containing buffer, i.e. the RIPA buffer, was also tested. RIPA contains a mixture of non-linear detergents, i.e. Triton X-100 (or NP-40) and deoxycholate. From the results of our preliminary tests, we decided to use beta-cyclodextrin instead of alpha-cyclodextrin. As beta-cyclodextrin is ten times less soluble in water than alpha-cyclodextrin, this limits the concentration of cyclodextrin that can be used in the assay and thus the amount of deterhent that can be tolerated. Nevertheless, as shown in Figure 5, an efficient protein assay can be devised, which tolerates up to 25µl of RIPA buffer per ml of assay. From the comparison of this beta-cyclodextrin-containing assay to the alpha-cyclodextrin-containing ones shown in figures 3 and 4, it can be seen that the slope of the curve is lesser in the case of the beta-cyclodextrin-containing assay, inducing a slightly less easy determination of the protein concentration.

Finally, to assess the performances of the cyclodextrin-containing assays with real complex samples, mammalian cells were extracted either with detergent-containing solutions or with solutions compatible with the standard Bradford assay (i.e. containing either urea and CHAPS or containing low concentrations of linear zwitterionic detergents. The concentrations of proteins in the various extracts was measured with the appropriate assays, either the standard Bradford for conditions i) and ii), the beta-cyclodextrin-containing assay for condition iii), and the alpha-cyclodextrin containing assays (5 mg/ml) for the conditions iv) to vi). The standard curves were made in the same buffer as the samples. Equal amounts, as determined from the various assays, were then loaded onto a SDS gel which was then stained with Coomassie Blue. the results, shown on Figure 6, show there is no gross difference between the various extracts, except for CTAB extracts. This can however be explained by the strong interference of CTAB with SDS electrophoresis. Indeed, in 2D protocols using cationic electrophoresis as the first dimension prior to SDS PAGE, the cationic detergent must be thoroughly removed prior to the SDS electrophoresis [24, 26].

Discussion

The method described above, named Cydex Blue (for CYcloDEXtrin- coomassie Blue) fulfills the requirements for a single step, easy to use protein assay compatible with both detergents of various classes and reducers. As the tolerance to detergents is limited by the complexing power of cyclodextrins, very diluted samples in concentrated detergents solutions can be difficult to assay. However this should not be a practical limit with most of the biological samples. An assay carried out with 5 mg/ml alpha-cyclodextrin tolerates up to 25μl of a 2% SDS sample and can easily read down to 2.5μg of protein, i.e. 0.1 mg/ml. If needed, an increase in the concentration of alpha-cyclodextrin can afford measurements of even more dilute samples. It can be seen from Figure 2 that concentration of up to 25 mg/ml alpha-cyclodextrin can be tolerated in the assay. From the 1:1 molecular ratio of cyclodextrin to detergent, 25mg alpha-cyclodextrin (MW 972) are able to complex 7mg SDS (MW 288), i.e. more than 300μl of a 2%SDS sample buffer, leading to an lower assay limit of 20μg protein per ml of sample buffer. Of course the presence of cyclodextrins alters the response of the assay, as does (but more weakly) the presence of the detergents. Consequently, for a better precision, the assay should be kept consistent, i.e. with a fixed amount of detergent-containing sample, a fixed concentration of cyclodextrin and a calibration curve made in the same conditions (detergent and cyclodextrin) as the samples to be assayed.

When linear detergents are used, alpha-cyclodextrin is the most convenient choice, although more expensive. It gives less background and a steeper response curve, leading to a better assay. Moreover, as it is highly soluble in water, its concentration i the assay mix can be tailored to the needs of the assay, e.g. for dilute samples.

Conversely, beta-cyclodextrin is less flexible to use, because of its lower solubility. It also gives more background and the background value is altered by the presence of detergents (Figure 1), which means that the assay must be very carefully controlled to be precise. This is probably due to the fact that beat -cyclodextrin probably complexes Coomassie Blue to some extent, leading a blue shift and thus to the background. However, detergents are better guests for beta-cyclodextrin than Coomassie Blue, so that they compete with Coomassie blue for beta-cyclodextrin. The expulsion of Coomassie Blue from the cyclodextrin could explain the decrease in background observed when detergents are added to a beta-cyclodextrin containing assay (Figure 1).

However, if rather concentrated samples are to be assayed, typically higher than 1mg protein per ml, then a beta-cyclodextrin-based assay could become economically attractive. In addition, beta-cyclodextrin offers a better scope in terms of compatible detergents, as it is able to bind linear as well as bulky detergents (e.g. Triton X-100).

In conclusion, the addition of cyclodextrins to a dye-binding, Bradford type protein assay represents a flexible tool to measure in a single step protein concentrations in the buffers commonly encountered in biochemistry (e.g. RIPA buffer, Laemmli type buffer but without tracking dye). The type and concentration of cyclodextrin to use can be tuned according to the composition and concentration of the samples, as outlined in the present paper.

Acknowledgements
TR thanks the CNRS, the CEA and the Université Grenoble Alpes for supporting this work.

Conflict of interest
The author declare no conflict of interest

References

[1] Romero-Calvo, I., Ocon, B., Martinez-Moya, P., Suarez, M. D., Zarzuelo, A., Martinez-Augustin, O., de Medina, F. S., *Anal Biochem* 2010, *401*, 318-320.

[2] Welinder, C., Ekblad, L., *J Proteome Res* 2011, *10*, 1416-1419.

[3] Lowry, O. H., Rosebrough, N. J., Farr, A. L., Randall, R. J., *J Biol Chem* 1951, *193*, 265-275.

[4] Smith, P. K., Krohn, R. I., Hermanson, G. T., Mallia, A. K., Gartner, F. H., Provenzano, M. D., Fujimoto, E. K., Goeke, N. M., Olson, B. J., Klenk, D. C., *Anal Biochem* 1985, *150*, 76-85.

[5] Bradford, M. M., *Anal Biochem* 1976, *72*, 248-254.

[6] Watanabe, N., Kamei, S., Ohkubo, A., Yamanaka, M., Ohsawa, S., Makino, K., Tokuda, K., *Clin Chem* 1986, *32*, 1551-1554.

[7] Fujita, Y., Mori, I., Kitano, S., *Chem Pharm Bull (Tokyo)* 1984, *32*, 4161-4164.

[8] Olson, B. J., Markwell, J., *Curr Protoc Protein Sci* 2007, *Chapter 3*, Unit 3 4.

[9] Hill, H. D., Straka, J. G., *Anal Biochem* 1988, *170*, 203-208.

[10] Bensadoun, A., Weinstein, D., *Anal Biochem* 1976, *70*, 241-250.

[11] Peterson, G. L., *Anal Biochem* 1977, *83*, 346-356.

[12] Sheffield, J. B., Graff, D., Li, H. P., *Anal Biochem* 1987, *166*, 49-54.

[13] Gracia, E., Fernandez-Belda, F., *Biochem Int* 1992, *27*, 725-733.

[14] Morcol, T., Subramanian, A., *Anal Biochem* 1999, *270*, 75-82.

[15] Ramagli, L. S., Rodriguez, L. V., *Electrophoresis* 1985, *6*, 559-563.

[16] Palepu, R., Reinsborough, V. C., *Can. J. Chem.-Rev. Can. Chim.* 1988, *66*, 325-328.

[17] Ohtani, Y., Irie, T., Uekama, K., Fukunaga, K., Pitha, J., *Eur J Biochem* 1989, *186*, 17-22.

[18] Atger, V. M., de la Llera Moya, M., Stoudt, G. W., Rodrigueza, W. V., Phillips, M. C., Rothblat, G. H., *J Clin Invest* 1997, *99*, 773-780.

[19] Chung, J. A., Wollack, J. W., Hovlid, M. L., Okesli, A., Chen, Y., Mueller, J. D., Distefano, M. D., Taton, T. A., *Anal Biochem* 2009, *386*, 1-8.

[20] Lin, C. E., Huang, H. C., Chen, H. W., *J Chromatogr A* 2001, *917*, 297-310.

[21] Yamamoto, E., Yamaguchi, S., Nagamune, T., *Anal Biochem* 2008, *381*, 273-275.

[22] Tastet, C., Lescuyer, P., Diemer, H., Luche, S., van Dorsselaer, A., Rabilloud, T., *Electrophoresis* 2003, *24*, 1787-1794.

[23] Neuhoff, V., Arold, N., Taube, D., Ehrhardt, W., *Electrophoresis* 1988, *9*, 255-262.

[24] Macfarlane, D. E., *Analytical Biochemistry* 1989, *176*, 457-463.

[25] Buxbaum, E., *Anal Biochem* 2003, *314*, 70-76.

[26] Navarre, C., Degand, H., Bennett, K. L., Crawford, J. S., Mortz, E., Boutry, M., *Proteomics* 2002, *2*, 1706-1714.

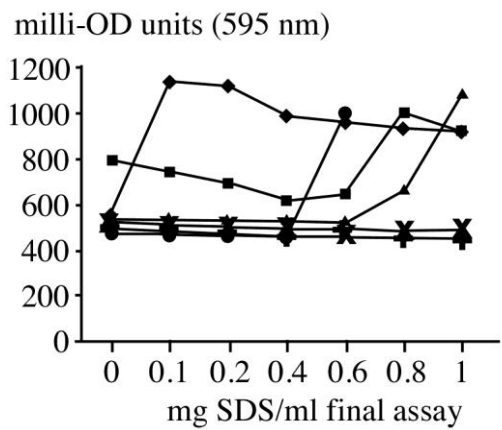
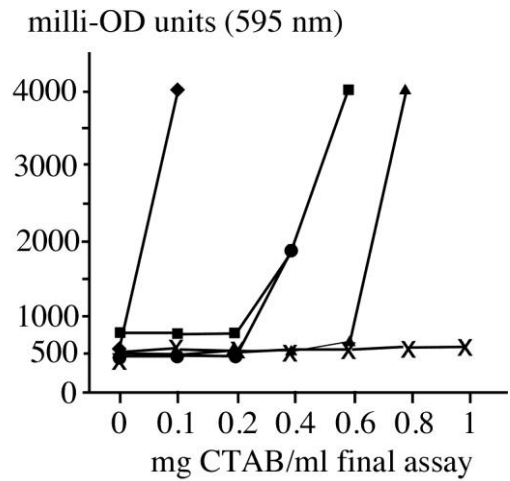

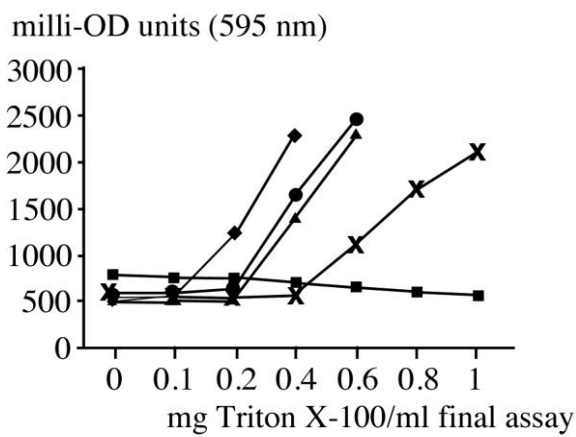

Figure 1: influence of the presence of cyclodextrins in a Bradford assay on the tolerance of the assay to various detergents. The zero was made with distilled water.

Panel 1A: SDS

diamonds: SDS without cyclodextrins

squares: + beta-cyclodextrin 2.5 mg/ml

circles: + alpha-cyclodextrin 2.5 mg/ml

triangles: + alpha-cyclodextrin 5 mg/ml

plus: + alpha-cyclodextrin 10 mg/ml

crosses: + alpha-cyclodextrin 25 mg/ml

Panel 1B: CTAB

diamonds: CTAB without cyclodextrins

squares: + beta-cyclodextrin 2.5 mg/ml

circles: + alpha-cyclodextrin 5 mg/ml

triangles: + alpha-cyclodextrin 10 mg/ml

crosses: + alpha-cyclodextrin 25 mg/ml

Panel 1C: Triton X-100

diamonds: Triton X-100 without cyclodextrins

squares: + beta-cyclodextrin 2.5 mg/ml

circles: + alpha-cyclodextrin 5 mg/ml

triangles: + alpha-cyclodextrin 10 mg/ml

crosses: + alpha-cyclodextrin 25 mg/ml

It can be seen that increasing amounts of alpha-cyclodextrin shift the point at which the detergent interference appears. Beta-cyclodextrin is more efficient than alpha-cyclodextrin for Triton X-100, and slightly more efficient for ionic detergents. Beta-cyclodextrin induces however a higher zero value and a less flat baseline when the amount if detergent is changed.

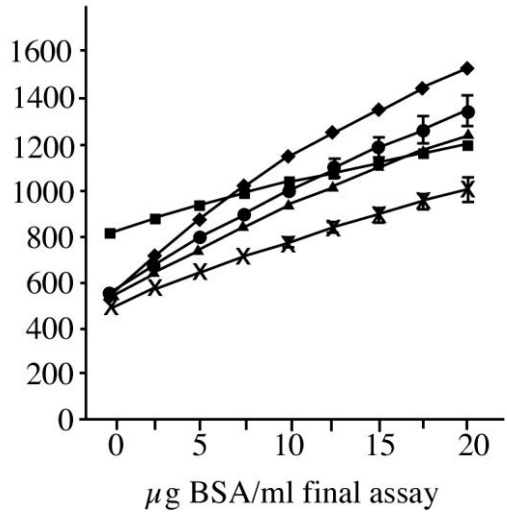

Figure 2: Influence of the presence of cyclodextrins on the response of a Bradford assay for BSA. Various amounts of cyclodextrins were added to the assay tube. The zero was made with distilled water.

diamonds: without cyclodextrins
squares: + beta-cyclodextrin 2.5 mg/ml
circles: + alpha-cyclodextrin 5 mg/ml
triangles: + alpha-cyclodextrin 10 mg/ml
crosses: + alpha-cyclodextrin 25 mg/ml

The slope of the curve decreases with increasing of alpha-cyclodextrin, and is also much flatter for beta-cyclodextrin.

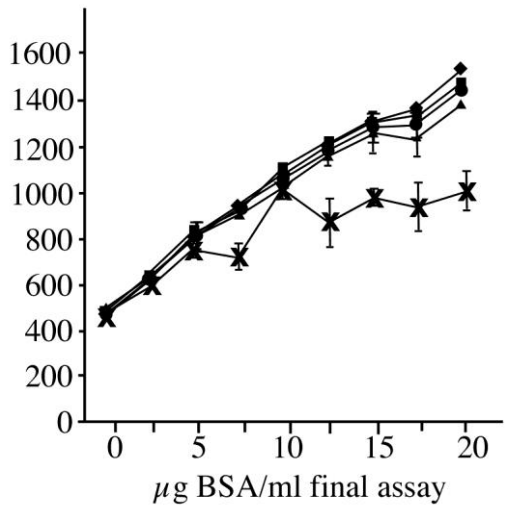 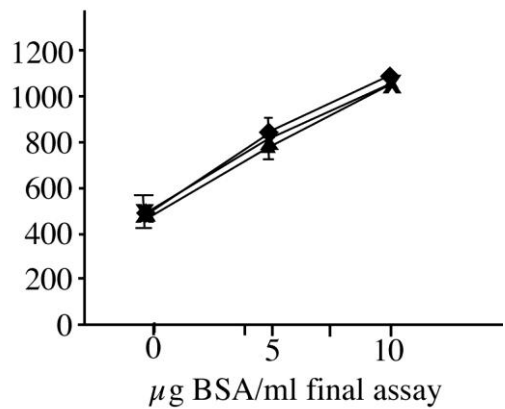

Figure 3: protein assay in SDS-containing buffer

The dose response curve of a Bradford type assay containing 5 mg/ml alpha-cyclodextrin for BSA in the presence of various amounts of SDS-containing buffer buffer was determined. The zero was made with distilled water. Unrepresented errors bars are smaller than the symbols size.

Panel 3A: influence of the amount of SDS
diamonds: without SDS
squares: With 10μl/ml SDS2% DTT 50mM
circles: With 20μl/ml SDS2% DTT 50mM
triangles: With 30μl/ml SDS2% DTT 50mM
crosses: With 40μl/ml SDS2% DTT 50mM

Panel 3B: influence of the reducer present in the SDS buffer
diamonds: 20 μl/ml SDS2%, betamercaptoethanol 5%
triangles: 20 μl/ml SDS2%, DTT 50mM
crosses: 20 μl/ml SDS2%, TCEP 10mM

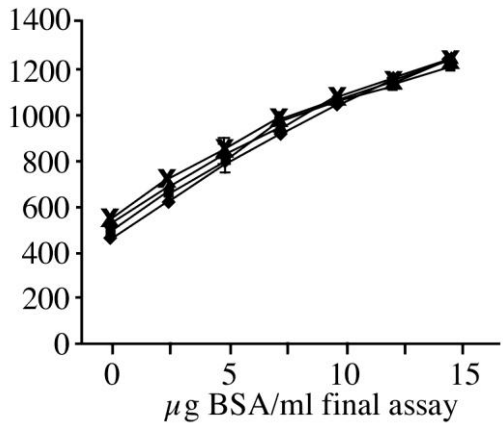 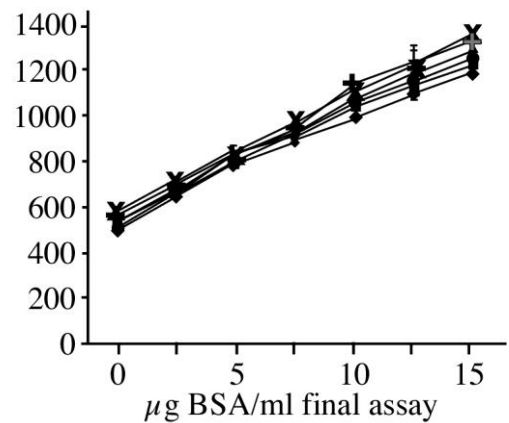

A  B

Figure 4: protein assay in CTAB-containing buffer

The dose response curve of a Bradford type assay containing 5 mg/ml alpha-cyclodextrin (Figure 4A) or 10 mg/ml alpha-cyclodextrin for BSA in the presence of various amounts of CTAB-containing buffer buffer was determined. The zero was made with distilled water. Unrepresented errors bars are smaller than the symbols size.

Panel 4A
diamonds: without CTAB
squares: With 10μl/ml CTAB2%, TCEP 10mM
triangles: With 20μl/ml CTAB2%, TCEP 10mM
crosses: With 25μl/ml CTAB2%, TCEP 10mM

Panel 4B
diamonds: without CTAB
squares: With 10μl/ml CTAB2%, TCEP 10mM
circles: With 20μl/ml CTAB2%, TCEP 10mM
triangles: With 30μl/ml CTAB2%, TCEP 10mM
plus: With 40μl/ml CTAB2%, TCEP 10mM
crosses: With 50μl/ml CTAB2%, TCEP 10mM

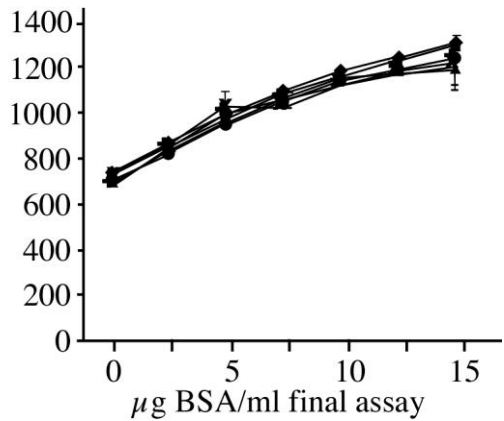

Figure 5: protein assay in RIPA buffer

The dose response curve of a Bradford type assay containing 2.5 mg/ml beta cyclodextrin for BSA in the presence of various amounts of RIPA buffer was determined. The zero was made with distilled water. Unrepresented errors bars are smaller than the symbols size.

diamonds: without RIPA Buffer
squares: With 5µl/ml RIPA Buffer
circles: With 10µl/ml RIPA Buffer
triangles: With 15µl/ml RIPA Buffer
plus: With 20µl/ml RIPA Buffer
crosses: With 25µl/ml RIPA Buffer

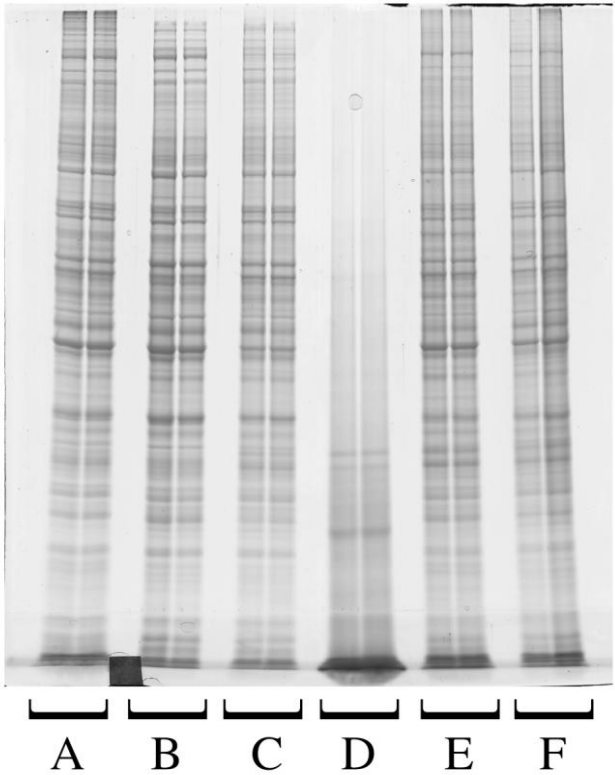

Figure 6: Verification of protein assay by SDS electrophoresis

proteins extracted from RAW264 cells under different condition were assayed by the corresponding Bradford-type assays. Conventional Bradford assay for the urea and SB 3-14 extracts, assay with 2.5 mg/ml beta-cyclodextrin for the RIPA extract, assay with 5mg/ml alpha-cyclodextrin for the SDS- or CTAB-containing extracts.

The experiment (extraction + assay) was carried out on two independent cell pellets for each condition. The extracts were diluted to 1mg protein per ml in SDS sample buffer, and 40µg of proteins were loaded for each sample on top of a 10% SDS-electrophoresis gel. After migration, the gels were stained with colloidal Coomassie Blue

A: urea/thiourea extract
B: Hepes/SB 3-14 extract
C: RIPA buffer extract
D: CTAB/TCEP extract
E: SDS/TCEP/pyruvate extract
F: SDS/DTT/Tris extract

Supplementary protocol

# Simplified starting protocol for the Cydex Blue assay

1) prepare a 5mg/ml cyclodextrin solution in hot water. Use preferentially alpha-cyclodextrin for samples containing linear detergents (e.g. SDS, CTAB, Brij) and beta cyclodextrin for sampel containing bulky detergents (Triton, bile salts, RIPA buffer). If in doubt, use beta-cyclodextrin

2) Prepare the standard curve: in a series of 1.5 or 2 ml tubes, add 300µl of water, 500µl of the cyclodextrin solution, 10µl of the sample buffer used for protein extraction, and various amounts (0-20µg) of BSA.

3) Prepare your samples: dilute your protein-containing samples in the corresponding sample buffer (a 1:10 starting dilution is often a good choice). in a series of 1.5 or 2 ml tubes, add 300µl of water, 500µl of the cyclodextrin solution, and finally 10µl of the diluted samples.

4) In all the tubes (standard curve and samples) add 200µl of 5x concentrated Bradford reagent (e.g. Bio-Rad catalog number #500-0006, but a home-made reagent can be prepared from the original recipe of Bradford). As the reagent is viscous, the use of a positive displacement pipette (e.g. Gilson Microman 250) is recommended. Mix immediately by inversion.

5) After 5 minutes at room temperature to reach a stable color, read the absorbance at 595 nm. If the samples are out of range, use a different dilution to reach a correct absorbance.